\documentclass[preprint,12pt]{elsarticle}
\usepackage{graphicx}

\usepackage{amssymb}

\journal{Physica D}

\newcommand{\be}{\begin{equation}}
\newcommand{\bea}{\begin{eqnarray}}
\newcommand{\ee}{\end{equation}}
\newcommand{\eea}{\end{eqnarray}}

\def\x{\mbox{{\bf x}}}

\def\g{\mbox{{\bf g}}}

\def\Y{\mbox{$\bf{Y}$}}
\def\X{\mbox{$\bf{X}$}}

\def\y{\mbox{$\bf{y}$}}

\def\p{\mbox{$\bf{p}$}}

\def\f{\mbox{$\bf{f}$}}
\def\F{\mbox{$\bf{F}$}}

\begin{document}

\begin{frontmatter}



\title{{\bf {\Large Self-Consistent Stochastic Model Errors\\ in Data Assimilation }}}


\author{Henry D. I Abarbanel}

\address{Department of Physics\\ and\\ Marine Physical Laboratory (Scripps Institution of Oceanography)\\ 
Center for Theoretical Biological Physics\\
University of California,
San Diego,\\ 
9500 Gilman Drive, Mailcode 0402,\\ 
La Jolla, CA 92093-0402   USA\\ 
habarbanel@ucsd.edu}

\begin{abstract}
In using data assimilation to import information from observations to estimate parameters and state variables of a model, one must assume a distribution for the noise in the measurements and in the model errors. Using the path integral formulation of data assimilation~\cite{abar2009}, we introduce the idea of self consistency of the distribution of stochastic model errors: the distribution of model errors from the path integral with observed data should be consistent with the assumption made in formulating the path integral. The path integral setting for data assimilation is discussed to provide the setting for the consistency test. Using two examples drawn from the 1996 Lorenz model, for $D = 100$ and for $D = 20$, we show how one can test for this consistency with essentially no additional effort than that expended in extracting answers to interesting questions from data assimilation itself.

\end{abstract}


\end{frontmatter}

\section{Introduction} 
Assimilating information from observed data to models of the observed system 
requires a formulation of the way in which the data communicates information to the model and a formulation of the manner in which the
model propagates states between observations. In the former, one must provide an approximation to the conditional mutual information
between the model state and the observation at each observation time. For the latter, one must provide
an approximation to the model of the dynamical processes along with the errors in the model. As one has some control over errors in measurements and models of the measurement, we focus here on errors in the models and how they are represented.

There are two general types of errors in models: (a) deterministic errors associated with dynamical processes absent from the model, and (b) stochastic errors associated with the resolution of the model as discretized in space and time or associated with noise that obscures perfect resolution. Unresolved processes at small space/time scales usually fall into the second category. One of the goals of data assimilation is to use information from measurements to estimate unknown constant parameters in a model, so we do not consider unknown values of fixed parameters as model errors.

In the next section we will be more precise about models and observations. For the discussion at this point, designate the D-dimensional state of the model at time $t_n$ as $\x(t_n) = \x(n)$. A dynamical model in discrete time (if partial differential equations underlie the observations, spatial variables have also been discretized) will give a rule for taking the state of the system $\x(n)$ to the state at a later time $t_{n+1}$, $\x(n+1)$, of the form $\g(\x(n+1),\x(n),\p_M) = 0$, where a set of fixed parameters for the model $\p_M$ is indicated.  If the connection between $\x(n)$ and $\x(n+1)$ is explicit, we would use $\g(\x(n),\x(n+1),\p_M) = \x(n+1) - \f(\x(n),\p_M)$ for the model error. In what follows we use these two expressions interchangeably.

Errors of type (a) above can be designated as corrections to a deterministic model by writing $\x(n+1) = \f(\x(n)) + \Delta \f(\x(n)$. This gives us little information on how the model error is to be represented as the correction $\Delta \f(\x(n))$ to the vector field $\f(\x(n),\p_m)$  is so general. Indeed we know of no algorithmic manner in which to select such errors in the model. 

Stochastic errors lead away from a deterministic model and require our focus on the probability of transition to $\x(n+1)$ given $\x(n)$: $P(\x(n+1)|\x(n))$. In a deterministic setting $P(\x(n+1)|\x(n)) = \delta^D(\x(n+1) - \f(\x(n),\p_M)) = \delta^D(\g(\x(n+1),\x(n),\p_M))$, and in the presence of stochastic model error the delta function is broadened. How it is broadened requires an assumption, and in this paper we propose a systematic manner to examine that assumption in the case where data from observations are available. Absent observations, any assumption about model errors is consistent, and, in fact, it prescribes the model itself.

In the next section we recall the general formulation of data assimilation in situations where there are noisy observations, model errors, and uncertainty in the state of the model when observations begin at $t_0$, $P(\x(0))$. This gives us a path integral over model states through an observation window~\cite{abar2009,alexander05,apte,restrepo}, yielding the conditional distribution of model states, from which expected state and parameter values can be evaluated.

Though their results were not stated as a path integral, the studies by~\cite{cox64,fried66,jaz} capture much of the structure we discuss here. Their focus was on iterative determinations of the joint conditional probability density $P(\X|\Y)$ and on maximizing that quantity. This seeks the mode of the distribution, while the path integral formulation allows evaluation of moments such as the expected value, covariances  about that, and marginal distributions of state and parameter values such as we explore here.

To utilize the general formulation one must make an assumption about the distribution of model errors in the transition probability $P(\x(n+1)|\x(n))$ wherein the model dynamics is placed. In the presence of observations, this assumption may or may not be correct, and one constructive way to examine the assumption is to compare moments of the model error $ME_a(n) = g_a(\x(n+1),\x(n))$ or $ME_a(n) = x_a(n+1) -f_a(\x(n));\;a=1,2,..,D; \;n=0,1, ...$ during a period of observations with one's assumed distribution. If the moments one evaluates are consistent with the assumed $P(\x(n+1)|\x(n))$, one's confidence in the overall data assimilation effort is supported. If not, another set of assumptions on how model errors are distributed must be examined. The required moments are themselves taken from the path integral, and consistency of the model errors is conditional on the observations and on one's knowledge or assumption on their distribution.

To illustrate these ideas, we examine `data' in a twin experiment~\cite{losch} using the D = 100 Lorenz96 model~\cite{lor96}. We add Gaussian noise to the `observations'. Because we generate the data in a twin experiment and use the data assimilation procedure to estimate the states of the model from sparse noisy observations, this exercise shows the procedure to be self-consistent when we have assumed a Gaussian broadening of the deterministic transition probabilities when sufficient observations are presented to the model.

We then examine another Lorenz96 model, this time with D = 20, and add noise distributed according to the gamma distribution to both the observations and to the dynamical equations. Then, on assuming these measurement and stochastic model error noise terms are Gaussian for purposes of assimilating data into the model, we show the Gaussian assumption is inconsistent.

In this paper we do not consider deterministic errors in model dynamics. As noted, the possibilities are too many, and we have no general or algorithmic manner in which to represent them. Perhaps one could examine a simple, overall bias term in this context, but we do not do so here. Further, if one finds, as we do in our second example, that the assumption about stochasticity in the data assimilation procedure is inconsistent with one's model and one's observed data, we do not propose a systematic remedy.

\section{General Formulation}

We begin by establishing a framework for assimilating information from observed data to a model of the observed system. To start we think of a physical system which is observed at times $t_n$ in the observation window $\{t_0, t_1, ..., t_m = T\}$,
and at each of these times we make $L$ observations $y_l(t_n) = y_l(n);\;l=1,2,...,L$. Independently of making the observations, we develop a model of the physical system from considerations of the physical or biophysical processes selected to be important in the development of the dynamics. If the observed physical processes are described by partial differential equations, we discretize both space and time to arrive at a discrete time map for the D-dimensional model state vector $x_a(t_n);\;a=1,2,...,D;\;n=0,1,...,m$. This model has $D$ degrees of freedom and consists of a rule taking the D-dimensional model state $\x(t_n) = \x(n)$ to the model state at time $t_{n+1}:\;x_a(t_{n+1}) = x_a(n+1) = f_a(\x(n),\p_M);\;a=1,2,...,D$ where $\p_M$ are time independent parameters in the model. 

The model should also specify how the state $\x(n)$ is related to the observations $y_l(n)$ through a set of $L$ observation functions $h_l(\x(n),\p_O)$ where $\p_O$ are fixed parameters in the observation functions. The total collection of parameters we will call $\p=\{\p_M,\p_O\}$. 

The goal of data assimilation is to use the information in the $\y(n)$, where $L < D$ in essentially all cases, and often $L << D$, to estimate both the parameters $\p$ and the $D - L$ unobserved states. If one does this over the observation window to achieve an estimation of the $\p$ and $\x(t_m = T)$, then one may use this information for characterizing the system state, at least according to the model, and for making predictions for $t > T$ using the dynamics prescribed through the model.

When the measurements $\Y(m) = \{\y(m),\y(m-1),...,\y(1),\y(0)\}$ are noisy, and the model has errors arising from resolution or missing terms, that is, physical processes not properly incorporated in the model, then the description of the outcome of using the model and of utilizing the information in the observations is in terms of a conditional probability distribution $P(\x(m)|\Y(m))$ for the probability of the state being at $\x(m)$ at the end of the observation window, given the sequence of measurements $\Y(m)$. 

In this stochastic setting such a probability distribution satisfies a Fokker-Planck equation~\cite{klwest} with familiar drift and diffusive terms, and when the observation noise and the model errors are not Gaussian, it also has an infinite number of additional terms. Since the Fokker-Planck equation is linear in the conditional probability, one would expect to find an integral representation of the desired probability distribution at time $t_m = T$ in terms of the probability distribution at $t_0:\; P(\x(0))$:
\be 
P(\x(m)|\Y(m)) = \int \prod_{n=0}^{m-1} d^Dx(n) \, K(\X(m),\Y(m)) P(\x(0)),
\label{intrep}
\ee
with $\X(m) = \{\x(m),\x(m-1),...,\x(1),\x(0)\}$.

Using identities on conditional information (Bayes' rule)~\cite{mackay} and the Chapman-Kolmogorov relation~\cite{pap}, true for Markov processes where $\x(n+1)$ depends only on $\x(n)$, as here, we derived~\cite{abar2009} a recursion relation between the conditional probability distribution $P(\x(n)|\Y(n))$ and $P(\x(n-1)|\Y(n-1))$:
\bea
P(\x(n)|\Y(n))&=&\frac{P(\x(n),\y(n)|Y(n-1))}{P(\x(n)|Y(n-1))\,P(\y(n)|\Y(n-1))}\,\nonumber \\
&&\int d^Dx(n-1) P(\x(n)|\x(n-1))P(\x(n-1)|\Y(n-1)) \nonumber \\
&=& \exp[CMI(\x(n),\y(n)|\Y(n-1)))] \,\nonumber \\
&& \int d^Dx(n-1) P(\x(n)|\x(n-1))P(\x(n-1)|\Y(n-1)),
\eea
where the mutual information~\cite{fano} between the D-dimensional state $\x(n)$ and the L-dimensional measurement $\y(n)$, conditional on earlier measurements $\Y(n-1)$, is
\be 
CMI(\x(n),\y(n)|\Y(n-1)) = \log\biggl\{\frac{P(\x(n),\y(n)|Y(n-1))}{P(\x(n)|Y(n-1))\,P(\y(n)|\Y(n-1))}\biggr\}.
\ee

As discussed by Fano~\cite{fano} this is to be thought of as a variable on the space of states $\x(n)$ and observations $\y(n)$. The average mutual information over this quantity has certain positivity properties~\cite{fano,meister}, though the CMI itself can be positive or negative.

Iterating this recursion relation from the end of the observation period at $t_m$ back to its beginning at $t_0$, we have the integral representation of the solution to the underlying Fokker-Planck equation as 
\bea 
P(\x(m)|\Y(m)) &=& \int \prod_{n=0}^{m-1} d^Dx(n) \exp \biggl\{\sum_{n=0}^m\,CMI(\x(n),\y(n)|\Y(n-1)) \nonumber \\
&+& \sum_{n=0}^{m-1} \log[P(\x(n+1)|\x(n))] + \log[P(\x(0))]\biggr \}.
\eea
This identifies the kernel $K(\X(m),\Y(m))$ in Equation (\ref{intrep}).
 
An arbitrary function $F(\X)$ on the $(m+1)D$ dimensional path $\X$ through the observation window has an expectation value
\bea 
E[F(\X)|\Y(m)] &=& <F(\X)> \nonumber \\
&=& \frac{\int \prod_{n=0}^m d^D\x(n)\,F(\X)\,\exp[-A_0(\X,\Y)]}{\int \prod_{n=0}^m d^D\x(n)\,\exp[-A_0(\X,\Y)]} \nonumber \\
&=& \frac{\int d\X \,F(\X)\,\exp[-A_0(\X,\Y)]}{\int d\X \,\exp[-A_0(\X,\Y)]},
\label{pimean}
\eea
in which the ``action'' is
\bea 
-A_0(\X(m), \Y(m))&=& \biggl\{\sum_{n=0}^m\,CMI(\x(n),\y(n)|\Y(n-1)) \nonumber \\
&+& \sum_{n=0}^{m-1} \log[P(\x(n+1)|\x(n))] + \log[P(\x(0))]\biggr \}.
\eea

To estimate the expected value of the model state during and at the end of the observation window $F(\X) = \X_{\alpha}$ or the covariance
about this expected value $<(X_{\alpha} - <X_{\alpha}>)(X_{\beta} - <X_{\beta}>)>$, with $\alpha,\beta$ including both the D model state components and the $(m+1)$ time indices, or any other $<F(\X)>$ of interest, we must perform the indicated integration in what may well be a quite high dimensional space. 
In general this cannot be accomplished analytically when the dynamics describing the transition $\x(n) \to \x(n+1)$ is nonlinear, so approximations must be made to both the conditional mutual information term in the action and to the transition matrix $P(\x(n+1)|\x(n))$ term describing the model dynamics. With these approximations, one must still perform the $(m+1)D$ dimensional integrals approximately. 

If the model were without errors and the resolution of the state were perfect, the transition matrix would be
\be 
P(\x(n+1)|\x(n)) = \delta^D(\x(n+1) - \f(\x(n),\p_M)) = \delta^D(\g(\x(n),\x(n+1),\p_M)),
\ee
and in our examples below we use
\be 
\g(\x(n),\x(n+1),\p_M) = -\x(n) + \x(n+1) + \frac{\Delta t}{2}\biggl\{\F(\x(n),\p_M) + \F(\x(n+1),\p_M)\biggr\},
\ee
where the ordinary differential equations for the model after spatial discretization is
\be 
\frac{d\x(t)}{dt} = \F(\x(t), \p_M),
\ee 
and the time step in the time discretization is $\Delta t$.

It is an unlikely situation to have no model errors except in problems where we prescribe the model, generate data using the model with a selected set of parameters and initial conditions $\x(0)$ and then utilize the path integral or other format for data assimilation to estimate those parameters and state variables at the end of observations $\x(t_m = T)$. This kind of `twin experiment'~\cite{losch} is always helpful in testing data assimilation methods.

We must make an approximation to $P(\x(n+1)|\x(n))$. If we assume the delta function is broadened by Gaussian noise or errors in resolution we can replace
\be 
\delta^D(\x(n+1) - \f(\x(n),\p_M)) \to (\frac{R_f}{2 \pi})^{D/2}\,\exp[-\frac{R_f}{2}\,\sum_{a=1}^D \sum_{n=0}^{m-1} (x_a(n+1) - f_a(\x(n)))^2],
\label{moderr}
\ee 
which becomes the delta function as the inverse resolution $R_f \to \infty$. We do not continue to explicitly show the dependence of the vector field $\f(\x)$ on the parameters.
There are many approximate forms of a delta function which become the delta function when the width of the approximate distribution goes to zero, and the Gaussian case, though in common use, is here as an illustration.

To use the general result Equation (\ref{pimean}) for numerical estimation of $<F(\X)>$, we also require an approximation for the conditional mutual information term in the action. Though not necessary, it is common to assume that the measurements $y_l(n)$ are independent of each other at time $n$ and independent of measurements at earlier times (however, see~\cite{hamill}). In this setting one may represent the conditional mutual information terms in the action as
\be 
-\frac{R_m}{2}\sum_{l=1}^L\,\sum_{n=0}^m (y_l(n) - h_l(\x(n)))^2,
\ee
when the measurement errors are assumed to be distributed as a Gaussian with variance $R_m^{-1}$.

With these approximations, the action becomes 
\bea 
A_0(\X(m), \Y(m)) &=& \frac{R_m}{2}\sum_{l=1}^L\,\sum_{n=0}^m (y_l(n) - h_l(\x(n)))^2 \nonumber \\
&+& \frac{R_f}{2}\,\sum_{a=1}^D \sum_{n=0}^{m-1} (x_a(n+1) - f_a(\x(n)))^2 \nonumber \\
&-& \log[P(\x(0)].
\label{stdaction}
\eea
Importantly, when the dynamics $\f(\x)$ is nonlinear, this is not quadratic in the state variables, so the integral, Equation (\ref{pimean}), requires numerical evaluation.

The goal of this paper is to examine the consistency of assumptions such as Equation (\ref{moderr}). If the Gaussian assumption in Equation (\ref{moderr}) or any other specific assumption about the stochastic model error is to be consistent, then the distribution of the model error for $a=1,2,..,D; \;n=0,1, ...$
\be
ME_a(n) = x_a(n+1) -f_a(\x(n)) = g_a(\x(n),\x(n+1))
\ee
resulting from performing the integral Equation (\ref{pimean}) with $F(\X(m)) = ME_a(n)$ should be the same as when evaluating the expectation of $F(\X)$ directly from the assumption on $P(\x(n+1)|\x(n))$. 

This means that $<F(\X)>$ evaluated from Equation (\ref{pimean}) should numerically be the same as
\be 
<F(\X)>_{ME} = \int d\X(m) \prod_{n=0}^{m-1} P(\x(n+1)|\x(n))\, F(\X(m)).
\ee
While demonstrating this would require comparing the quantities  $<F(\X)>$ and $<F(\X)>_{ME}$ for arbitrary functions or, perhaps for all moments assuming they exist, we cannot do that in practice, so we examine a much more limited set of comparisons.

The reader will have realized that the equality of $<F(\X)>$ and $<F(\X)>_{ME}$ is equivalent to
\be 
\sum_{n=0}^m\,CMI(\x(n),\y(n)|\Y(n-1)) = 0,
\ee
and is essentially a test of how well the model, along with assumptions of representations of its stochastic errors, performs in matching the model output with the observations $\Y(m)$. 

This is also essentially the same as saying the model output $h_l(\x(n))$ at each measurement time $t_n$ is synchronized to the measurements $y_l(n)$ at those times. In the case of independent measurements with a Gaussian distribution of measurement errors Equation (\ref{moderr}), it also represents maximal information transfer from the observed data to the model.
In this spirit, under the assumption that noise in the measurements is additive, we could also examine the question of whether all aspects of the observation error $OE_l(n) = y_l(n) - h_l(\x(n));\;l=1,2,...,L;\;n=0,1,...,m$ are such that $<F(OE_l(n))> = F(0)$. In the remainder of this paper, we focus on considerations of the stochastic model error, recognizing its equivalence to the accuracy with which observation errors are essentially zero.

\section{A First Example from the Lorenz96 Model with D = 100}

The 1996 model of Lorenz~\cite{lor96} has D degrees of freedom $w_a(t);\;a=1,2,...,D$ in a periodic sequence satisfying the differential equations
\be
\frac{dw_a(t)}{dt} = w_{a-1}(t)(w_{a+1}(t) - w_{a-2}(t)) - w_a(t) + f,
\label{lor96eqn}
\ee
with $w_{-1}(t) = w_{D-1}(t), w_0(t) = w_{D}(t)$, and $w_{D+1}(t) = w_1(t),$ with some choice of $w_a(0)$. The `forcing constant' $f$ is a control parameter for the bifurcations of the solutions $w_a(t)$ of these equations, and when $f \approx 8$ or larger, the solutions exhibit chaotic behavior.

In this example, we selected $D = 100$ and $f = 8.17$ and generated a set of `data' $y_l(t_n)$ for $t_{n+1} - t_n = 0.01$ and $L = 40$. $L = 40$ was chosen from considerations of the smoothness of the action for this problem as a function of the number of measurements or equivalently from the number of positive conditional Lyapunov exponents on the synchronization manifold $x_l(n) \approx y_l(n)$ which, for the Lorenz96 problem, was determined to be $L \approx 0.4D$~\cite{kost1,kost2,quinn10}. The measurement function is the unit operator in the present example $h_l(\x(n)) = x_l(n)$. We used the Gaussian assumption of stochastic noise to approximate the action in the data assimilation path integral, so we adopted Equation (\ref{stdaction}) for the evaluation of $<F(\X)>$ for functions on the path $F(\X)$.

We generated `data' solving Equation (\ref{lor96eqn}) for a time step $\Delta t = 0.01$ and evaluated $y_l(n);\;l=0,1,...,D-1$ for $t_{n+1} - t_n = 0.05$, namely, every fifth time step in the development of the dynamics yields observations $\y(n)$. Data for 
\bea
&&l = \{0, 2, 5, 7, 10, 12, 15, 17, 20, 22, 25, 27, 30, 32, 35, 37, 40, 42, 45, \nonumber \\
&&47, 50, 52, 55, 57, 60, 62, 65, 67, 70, 72, 75, 77, 80, 82, 85, 87, 90, 92, 95, 97\} \nonumber
\eea
were presented to the model for an observation window $0 \le t \le 4$. No observations were presented in a prediction window $[4,6]$. The integrals involved were of dimension $(m+1)D = 60100$, and we evaluated them approximately using a standard Metropolis-Hastings Monte Carlo method~\cite{quinn10,neal,hast}. In this method we selected starting paths at random and performed 35,000 initialization path selections before recording statistics on $<F(\X)>$ for another 310,000 paths. The recorded paths were broken into blocks of 100 with 3100 blocks evaluated to approximate the integrals. These calculations were performed on a standard 2.5 GHz CPU. They were subsequently performed on a NVIDIA GPU with a speedup of about 75-100 in each case.

In the calculations we averaged over each block of 100 paths to achieve $N_{AP} =3100$ accepted paths $\X^{(j)}$ of length 60101 each, including the parameter $f$. With these paths we approximated $<F(\X)>$ as
\be 
<F(\X)> \approx \frac{1}{N_{AP}}\sum_{j=1}^{N_{AP}} F(\X^{(j)}),
\ee
as the accepted paths were distributed according to $\exp[-A_0(\X,\Y)]$ by the Metropolis-Hastings procedure. The expected error in this approximation to $<F(\X)>$ is a few parts in $10^{-2}$.

We evaluated the expected values of $F(\X)$ from the set $\{x_a(n)^q, ME_a(n)\}$ for $q = 1,2,3,4$ allowing us to estimate the expected mean path, the RMS variation about that path as well as the skewness and kurtosis about that mean. The latter allows us to examine whether the common assumption that the integrals involved are approximately Gaussian is correct. 
From the collection of accepted paths we are also able to estimate the marginal distributions of any element of the path, and in particular we were interested in the distribution of $ME_a(n)$.  If the assumption made in formulating the action Equation (\ref{stdaction}) that the stochastic model errors are distributed as a Gaussian, then we expect the mean of $ME_a(n)$ for any index $a$ and any time $n$ to be zero. We expect the RMS error variation about this mean to satisfy $RMS(ME_a(n)) \sqrt{R_f} = 1$ and the distribution $P(ME_a(n))$ for any $a$ and $n$ to be Gaussian.

When we generated our `data' by solving the Lorenz96 D = 100 equations, we added Gaussian noise with a signal to noise ratio of about 23 dB to the clean signal. This translates to $R_m \approx 8.0$, and we used that value in our Monte Carlo integrations. We also selected $R_f = 100$ as our experience with these methods~\cite{quinn10} suggests that $R_f \approx 10 R_m$ gives a sufficiently large $R_f$ that the imposition of the approximate equations of motion is accurate.

As examples of the outcome of these calculations, we report that for $ME_{76}(81)$ (chosen at random) the mean value was $9.9 \times 10^{-3}$, $RMS(ME_{76}(81))\,\sqrt{R_f} = 0.97$, and the skewness and kurtosis of this variable were smaller than 0.01 in magnitude. 

\begin{figure}[ht]
\includegraphics[width= 3.5in]{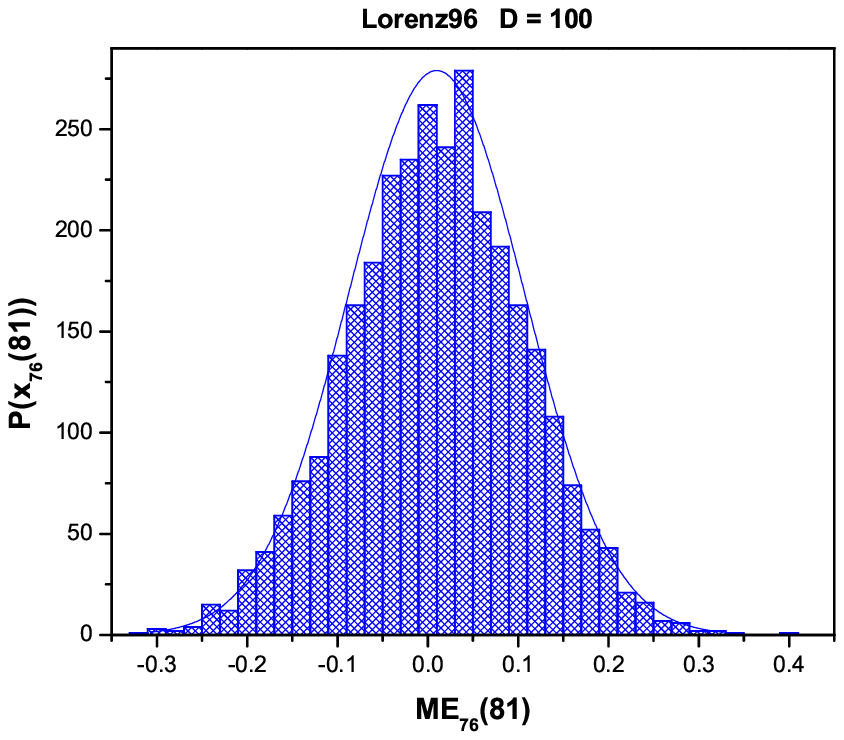}
\includegraphics[width= 3.5in]{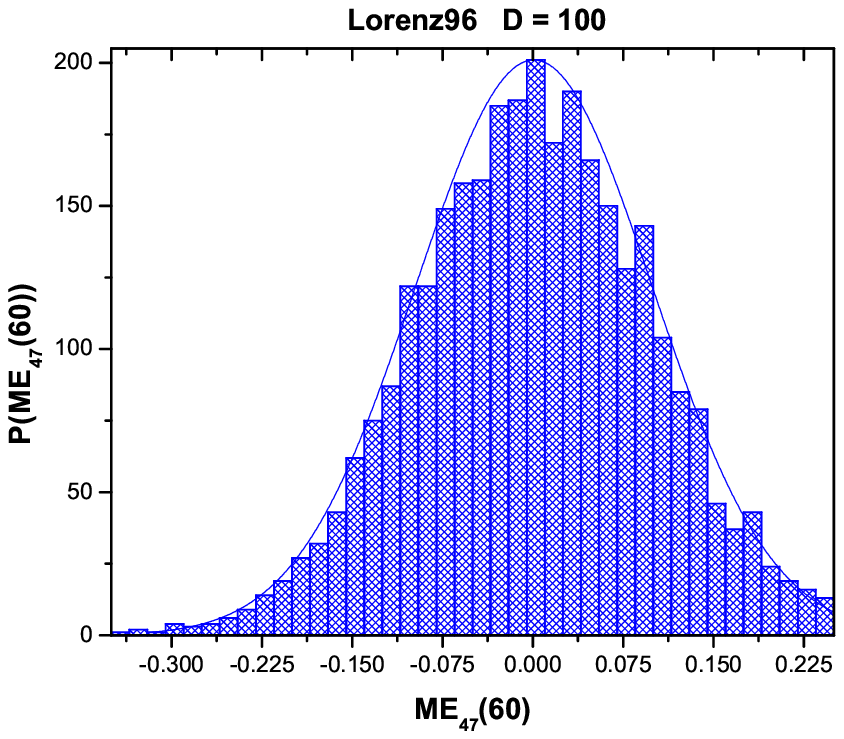}
\caption{{\bf Left} Distribution of the stochastic model error $ME_{76}(81)$ from the D = 100 Lorenz96 model compared to a best fit Gaussian distribution. {\bf Right} Distribution of the stochastic model error $ME_{47}(60)$ from the D = 100 Lorenz96 model compared to a best fit Gaussian distribution.}
\label{x76_81histoBnormal}
\end{figure}

Figure (\ref{x76_81histoBnormal}) {\bf Left} shows the distribution of values of $ME_{76}(81)$ over the 3100 paths along with the best fit Gaussian to that distribution. As $RMS(ME_{76}(81))\,\sqrt{R_f} = 0.97$, it is clear to the eye that this distribution of this model error is consistent with the assumed stochastic model error distribution. Figure (\ref{x76_81histoBnormal}) {\bf Right} also shows the distribution of $ME_{47}(60)$ from the same set of calculations compared again to the best fit Gaussian distribution. In this case the mean value of $ME_{47}(60)$ was $8.8 \times 10^{-4}$ and $RMS(ME_{47}(60))\,\sqrt{R_f} = 0.9899$; again the skewness and kurtosis are quite small as would be expected for a nearly Gaussian distribution. We can say for this case also that the output distribution of stochastic model errors is consistent with the assumed distribution.

In Figure (2) {\bf Left} we show the expected value of the observed variable $x_{47}(t)$ and its RMS error through the observation window $[0,4]$ and into the prediction region $[4,6]$. In Figure (2) {\bf Right} we show the expected value of the unobserved variable $x_{76}(t)$ and its RMS error through the observation window $[0,4]$ and into the prediction region $[4,6]$.  Figure (3) shows the skewness and kurtosis for the observed variable $x_{47}(t)$ through the window $[0,6]$. We see that the skewness and kurtosis are small in the observation window and then grow substantially after observations are terminated. This is consistent with the chaotic orbits of the model. Figure (4) displays the skewness and kurtosis of the unobserved variable $x_{76}(t)$, and we see that the values both within and without the observation window are larger than for the observed variable.

We noted above that the consistency of the assumed and the calculated distribution is equivalent to the precision of the equality of model output as estimated through the path integral and the observations presented to the model.

\section{A Second Example from the Lorenz96 Model with D = 20}

For a second example we again use the Lorenz96 model, Equation (\ref{lor96eqn}) but now with D = 20. We selected $f = 7.93$; again a value leading to chaotic orbits. We introduced noise into the dynamics and added noise to the observations presented to the model $y_l(n)$ for $L = 8$ and for
\be
l = \{0, 2, 5, 7, 9, 11, 17, 18\}.
\ee
The noise was taken from the Gamma distribution~\cite{feller}
\be 
P_{\Gamma}(x) = \frac{x^{a-1}e^{-x}}{\Gamma(a)},
\ee 
and added to each component of the model and to the `data' generated by the model as
\be 
\mbox{scale}(P_{\Gamma}(x) - a),
\ee
noting that
\be 
\int_0^{\infty} dx\,P_{\Gamma}(x)\,x = a.
\ee
We selected $a = 7$ and $\mbox{scale} = 0.05$ in the dynamical equations and $\mbox{scale} =0.205$ in the additive noise in the data $y_l(n)$. Again we chose $R_m = 8$ and $R_f = 100$ using $m = 410$ with $\Delta t = 0.02$ in the integration to produce the data. The path integrals were thus of dimension 8220, and we used 25,000 initialization Monte Carlo accept/eject steps to begin followed by 271,000 steps where statistics were recorded.

Selecting, again randomly, the model error $ME_{8}(91)$, we display the distribution of this model error in Figure (\ref{me11_104histonormal}) {\bf Right} along with a best fit Gaussian distribution. For this model error term, the mean value was -0.064, while
$RMS(ME_8(91)) \sqrt{R_f} = 0.609$ which is quite different from the unity required for consistency with the assumption of Gaussian broadening of the deterministic $P(\x(n+1)|\x(n))$ assumed in the path integral. One can see from Figure (\ref{me11_104histonormal})  that the computed distribution $P(ME_8(91))$ is significantly narrower than a Gaussian. The same calculation but for the model error term $ME_{11}(104)$ is shown in Figure (\ref{me11_104histonormal}) {\bf Left} again along with a best fit Gaussian. For this distribution the mean was $7 \times 10^{-3}$ and $RMS(ME_{11}(104)) \sqrt{R_f} = 0.64$. One can conclude that the assumption of Gaussian broadening of the transition probability used in the path integral and of Gaussian additive noise using in the conditional mutual information term of the action are not consistent with the data and the noisy model.

If we examine Figure (\ref{x17estandRMSD20}) where the expected value of the observed model variable $x_{17}(t)$ through the observation window $[0,8.2]$ is displayed along with the calculated RMS variation about that expected value and with the observed noisy `data' points presented to the model, we see sizeable regions where the estimated value $<x_{17}>(t)$ deviates from the observations. In Figure (\ref{x17skewnessD20}) one sees very large values of the skewness and kurtosis within the observation window for the {\tt observed} model variable $x_{17}(t)$ showing the departure of its distribution from a Gaussian. Figure (\ref{x14skewnessD20}) shows the same features for the {\tt unobserved} model dynamical variable $x_{14}(t)$.

Certainly more precise statistical tests can be made to determine the deviation of the distribution of the model errors selected here from Gaussians. As Gaussians were assumed in formulating the path integral as described above, we can conclude with confidence that the Gaussian assumption is inconsistent with the data. Of course, we built this into our `twin experiment' calculation, so it is perhaps reassuring that we are able to detect this inconsistency with essentially no more effort than already required in evaluating the data assimilation path integral for our other purposes: estimation of parameters and states within and at
the end of an observation period, prediction of the estimated states and their RMS errors beyond the observation window, ....
Once we have the accepted paths $\X^{(j)}$, evaluation of any $F(\X^{(j)})$ is quite straightforward.

In the more interesting situation of data from field or laboratory observations, we may make the same calculations of our assumed stochastic model error terms and check equally straightforwardly for the consistency of these assumptions.

\section{Discussion}

In assimilating information from measurements to a model of the observed system when the data are noisy, the models have error, and the state of the model system uncertain when measurements begin, one must make assumptions both about the way to represent stochastic model errors and noisy information transfers from the data. The latter are often known through knowledge of the sensors and the environmental noise during measurements. Model errors can be structural and deterministic arising from physical processes unaccounted for in developing the model or they can be stochastic representing limits on the spatial or temporal resolution of the model or environmental noise representing fluctuations on any scale not dynamically treated in the model. The stochastic model errors broaden the manner in which the dynamics enters data assimilation. In the deterministic case the transition probability to go from $\x(t_n) = \x(n)$ to $\x(t_{n+1}) = \x(n+1)$ is $P(\x(n+1)|\x(n)) = \delta^D(\x(n+1) - \f(\x(n)))$. This is broadened in the stochastic data assimilation ( or ensemble data assimilation) task, and an assumption on how this is represented must be made.

By comparing properties of the stochastic model errors $ME_a(n) = x_a(n+1) - f_a(\x(n))$ in the assumed distribution for $P(\x(n+1)|\x(n))$ and the properties emerging from the data assimilation procedure, we can test for the consistency of the assumptions about the stochastic model errors.

Using two variants of the Lorenz96 model, we examined a case where there was demonstrable consistency of the outcome of data assimilation and assumptions about the distribution of model errors, and we reported on another example where there was not such consistency. If the distribution of model errors is consistent between these two situations, this provides confidence in the precise formulation of the data assimilation tasks. Similarly, when that consistency is absent, confidence is lost. We do not provide a remedy in the case of inconsistency.

To execute this consistency test one requires basically the same numerical evaluations as in performing the overall data assimilation effort using the path integral formulation of the problem~\cite{abar2009}, so carrying out the consistency check is computationally quite inexpensive.

We do not give a remedy to finding inconsistency in the assumed and determined stochastic model error. However, in the dynamical parameter and state estimation procedure~\cite{kost1} we found that adding a control term $u(t)(y(t) - x(t))$ to the dynamics and a quadratic term in the `control' $u(t)$ to the equivalent of the action, we could extract indications of where in time model errors might occur and require compensation by the external forcing $u(t)$. We have not pursued the possibility here, but it seems attractive to consider adding such controls to both the model error terms and the mutual information terms of the action to trace the locations of errors in the model in phase space, though the temporal location of a requirement for large $u(t)$. If discretized space is part of the physical setting, that too could be traced in this manner.
\clearpage 
\section*{Acknowledgements}
The idea of examining model errors in data assimilation problems arose in discussions with Mike Fisher and Yanick Tremolet at the European Centre for Medium Range Weather Forecasts, and I thank them for suggesting the interest in the problems. Support from the US Department of Energy (Grant DE-SC0002349 ) and the National Science Foundation (Grants  IOS-0905076 and PHY-0961153) are gratefully acknowledged. Partial support from the NSF sponsored Center for Theoretical Biological Physics is also appreciated. Jack Quinn developed the Monte Carlo code for the data assimilation path integral~\cite{quinn10}.

\begin{figure}[ht]
\includegraphics[width= 3.5in]{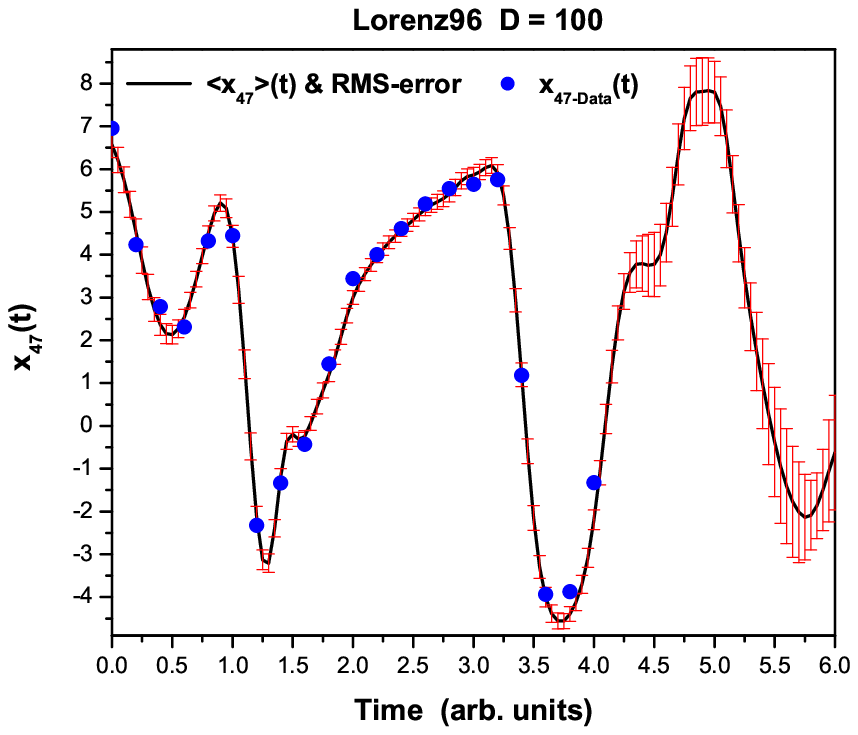}
\includegraphics[width= 3.5in]{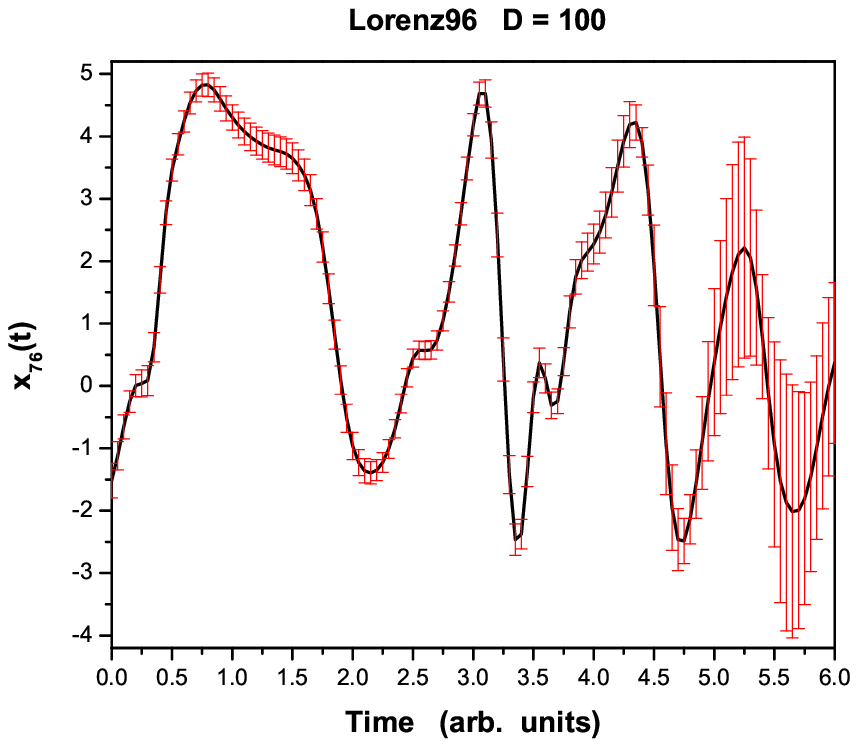}
\caption{{\bf Left} Using observed data (blue circles) in the observation window $[0,4]$, the expected
value of the (observed) variable $x_{47}(t)$ (black line) was estimated using the path integral discussed in the text. The RMS error (red error bars) around this expected value was also estimated. Using the estimated value of all 100 state variables (40 observed and 60 unobserved) along with the estimated parameter, predictions including RMS errors were made into $[4,6]$ as shown. Since the Lorenz96 D = 100 model is chaotic for the chosen forcing, the error bars grow in the prediction interval. {\bf Right} Using observed data in the observation window $[0,4]$, the expected
value of the (unobserved) variable $x_{76}(t)$ (black line) was estimated using the path integral discussed in the text. The RMS error (red error bars) around this expected value was also estimated. Using the estimated value of all 100 state variables (40 observed and 60 unobserved) along with the estimated parameter, predictions including RMS errors were made into $[4,6]$ as shown. Since the Lorenz96 D = 100 model is chaotic for the chosen forcing, the error bars grow in the prediction interval.}
\label{x47andrms}
\end{figure}

\begin{figure}[ht]
\includegraphics[width= 3.5in]{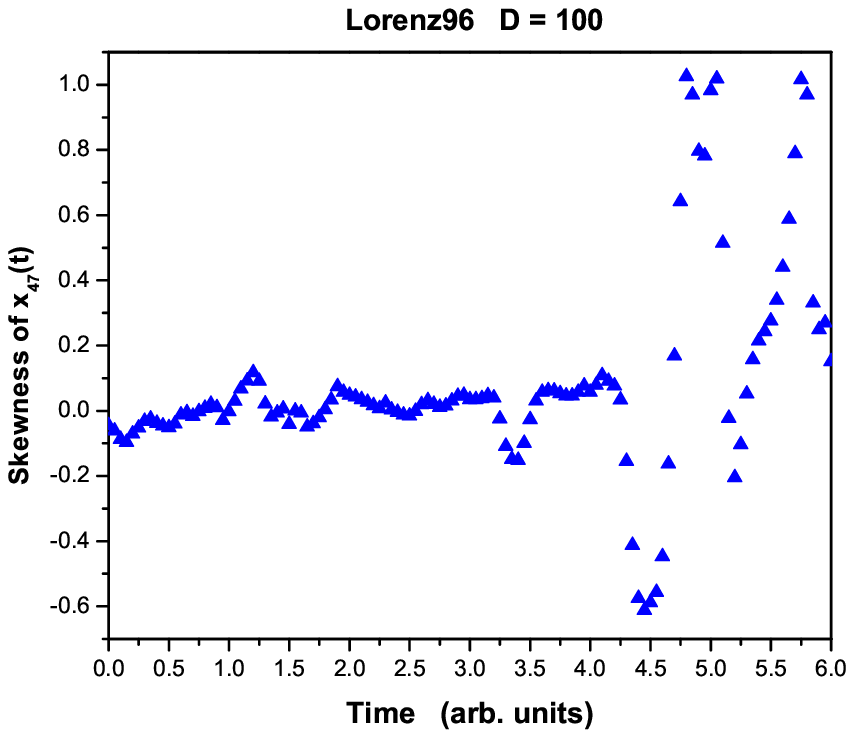}
\includegraphics[width= 3.5in]{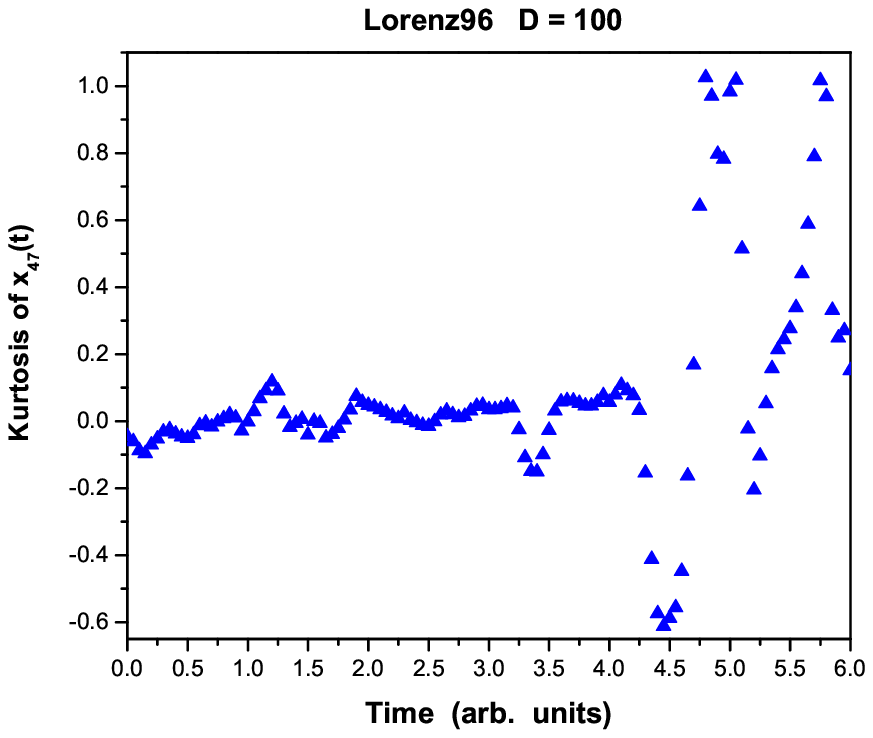}
\caption{{\bf Left} Skewness of the state variable $x_{47}(t)$, one of the observed quantities presented to the model though the data assimilation path integral. During the observation period the skewness remains small suggesting the state distribution may be approximately Gaussian in this interval. The skewness grows rapidly when observations no longer are available to guide the trajectory of the model and it moves rapidly to its attractor which is comprised of points in 100 dimensional state space which are not distributed as a Gaussian. {\bf Right} Kurtosis of the state variable $x_{47}(t)$, one of the observed quantities presented to the model though the data assimilation path integral. During the observation period the kurtosis remains small suggesting the state distribution may be approximately Gaussian in this interval. The kurtosis grows rapidly when observations no longer are available to guide the trajectory of the model and it moves rapidly to its attractor which is comprised of points in 100 dimensional state space which are not distributed as a Gaussian.}
\label{x47skewness}
\end{figure}



\begin{figure}[ht]
\includegraphics[width= 3.5in]{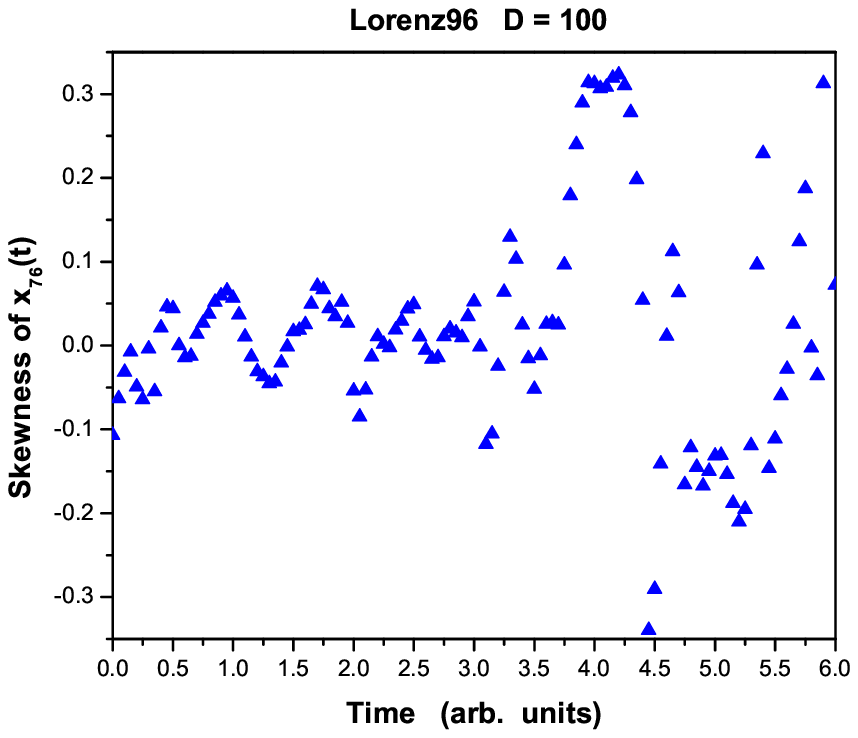}
\includegraphics[width= 3.5in]{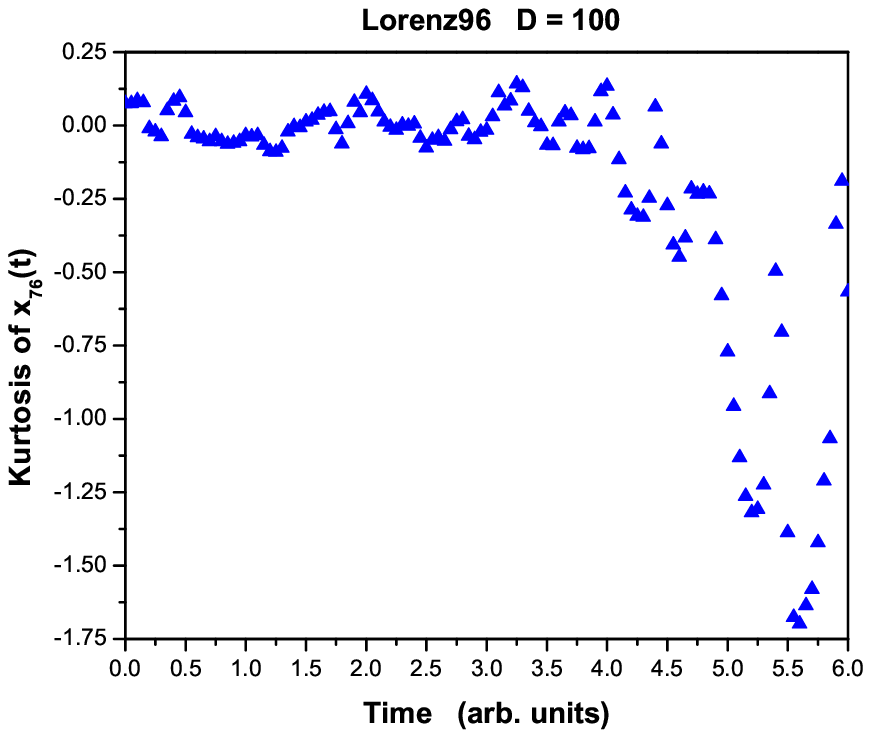}
\caption{{\bf Left} Skewness of the state variable $x_{76}(t)$, one of the unobserved quantities not presented to the model though the data assimilation path integral. During the observation period the skewness remains small suggesting the state distribution may be approximately Gaussian in this interval. The skewness grows rapidly when observations no longer are available to guide the trajectory of the model and it moves rapidly to its attractor which is comprised of points in 100 dimensional state space which are not distributed as a Gaussian. {\bf Right} Kurtosis of the state variable $x_{76}(t)$, one of the unobserved quantities not presented to the model though the data assimilation path integral. During the observation period the kurtosis remains small suggesting the state distribution may be approximately Gaussian in this interval. The kurtosis grows rapidly when observations no longer are available to guide the trajectory of the model and it moves rapidly to its attractor which is comprised of points in 100 dimensional state space which are not distributed as a Gaussian.}
\label{x76skewness}
\end{figure}


\begin{figure}[ht]
\includegraphics[width= 3.5in]{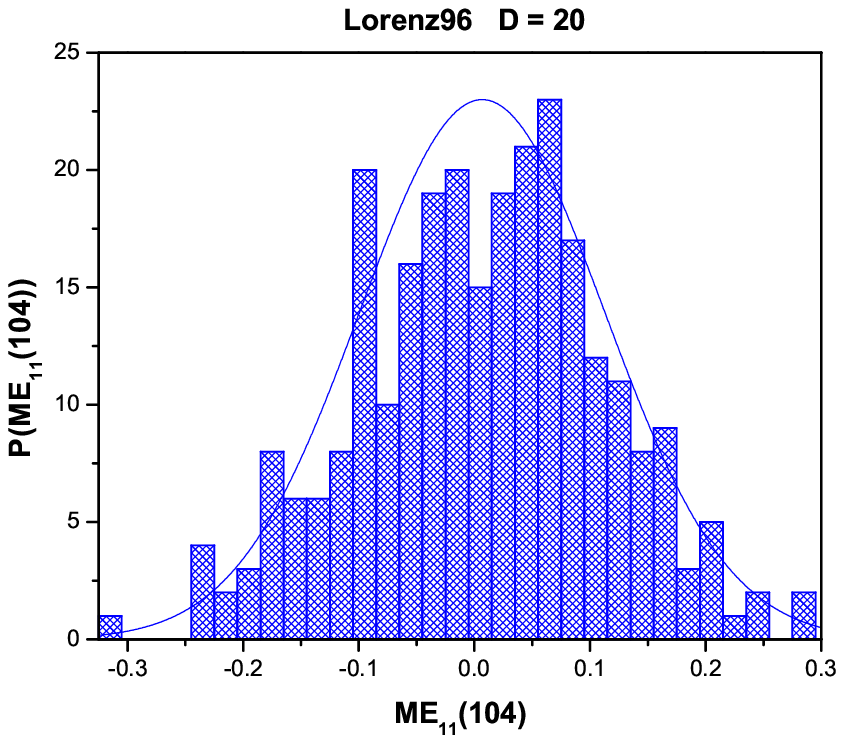}
\includegraphics[width= 3.5in]{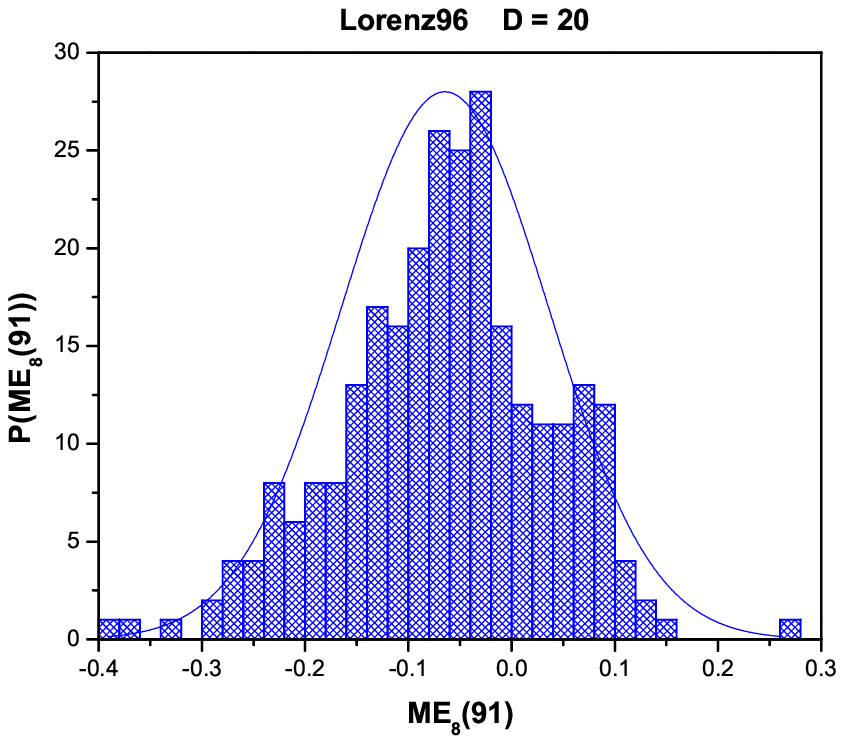}
\caption{{\bf Left} Distribution of the stochastic model error $ME_{11}(104)$ from the D = 20 Lorenz96 model compared to a best fit Gaussian distribution. {\bf Right} Distribution of the stochastic model error $ME_{8}(91)$ from the D = 20 Lorenz96 model compared to a best fit Gaussian distribution.}
\label{me11_104histonormal}
\end{figure}


\begin{figure}[ht]
\includegraphics[width= 5.4in]{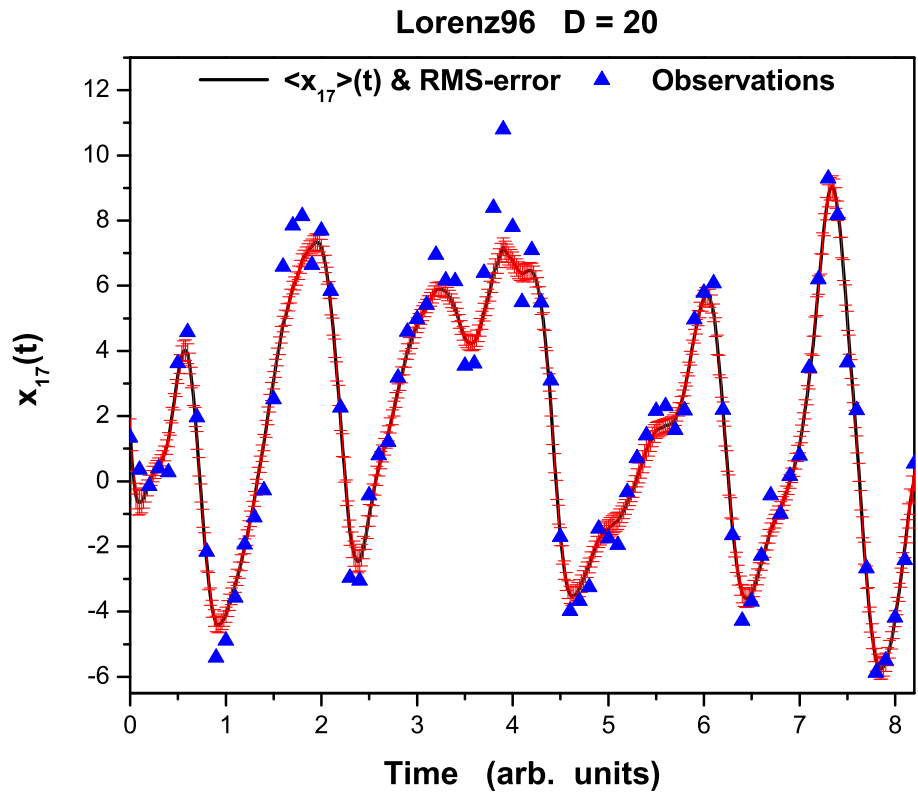}
\caption{Using observed data (blue triangles) in the observation window $[0,8.2]$, the expected
value of the (observed) variable $x_{17}(t)$ (black line) was estimated using the path integral discussed in the text. The RMS error (red error bars) around this expected value was also estimated.}
\label{x17estandRMSD20}
\end{figure}

\begin{figure}[ht]
\includegraphics[width= 3.5in]{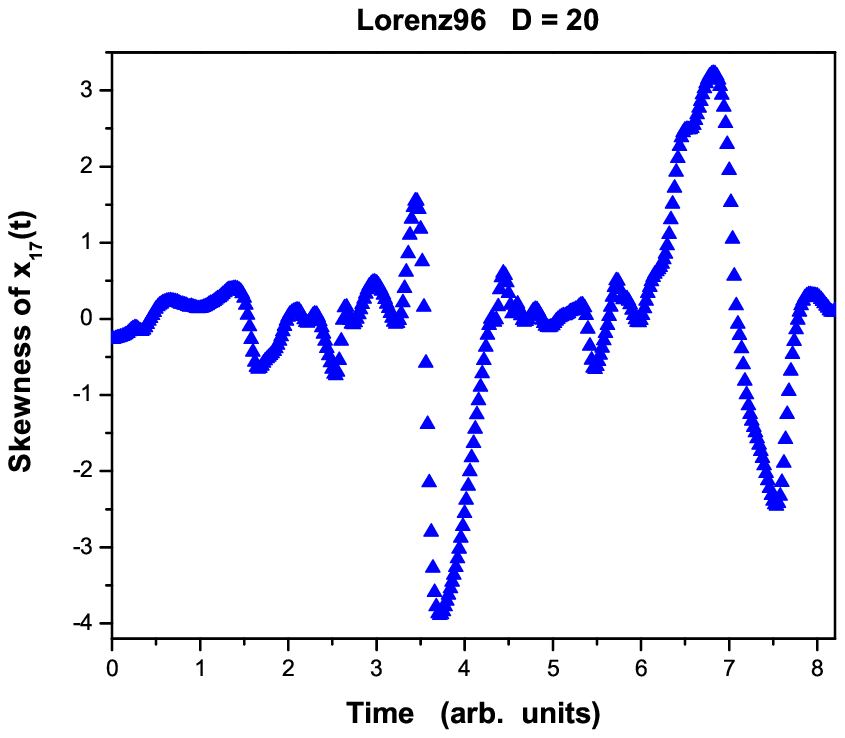}
\includegraphics[width= 3.5in]{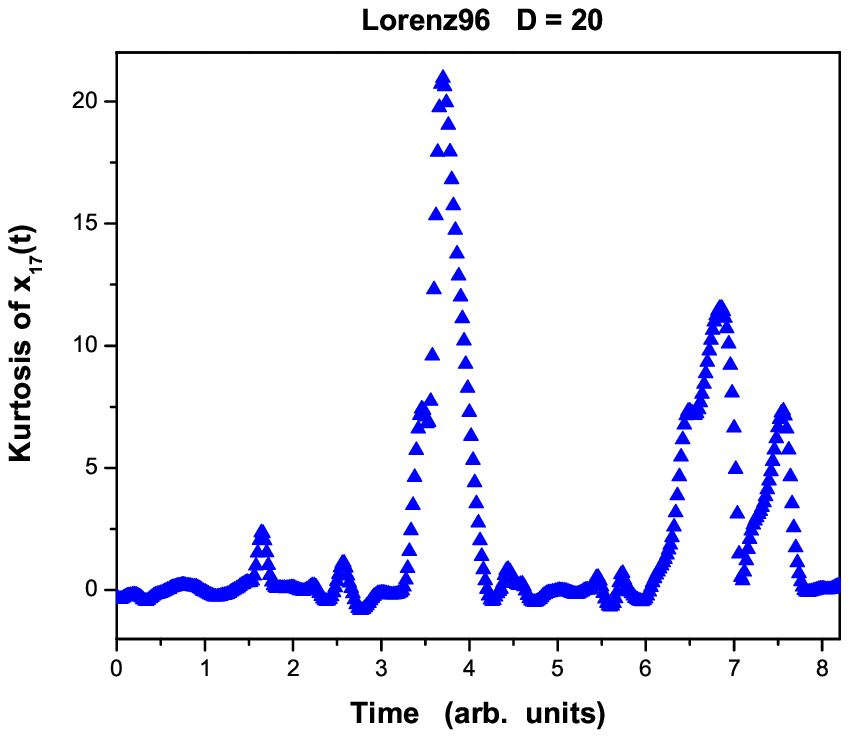}
\caption{{\bf Left} Skewness of the state variable $x_{17}(t)$, one of the observed quantities presented to the model though the data assimilation path integral. {\bf Right} Kurtosis of the state variable $x_{17}(t)$, one of the observed quantities presented to the model though the data assimilation path integral.}
\label{x17skewnessD20}
\end{figure}


\begin{figure}[ht]
\includegraphics[width= 3.5in]{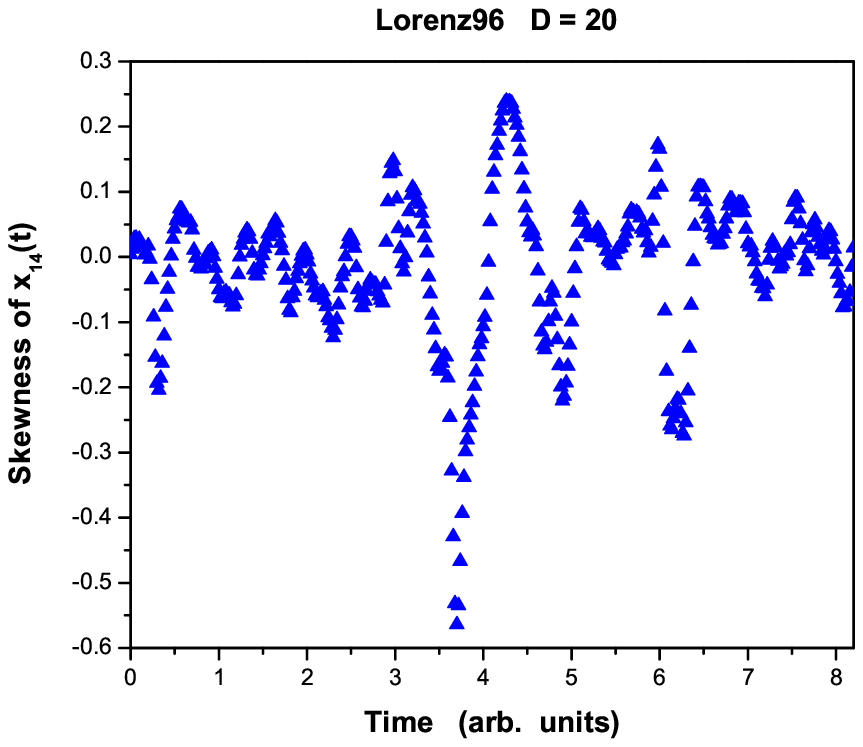}
\includegraphics[width= 3.5in]{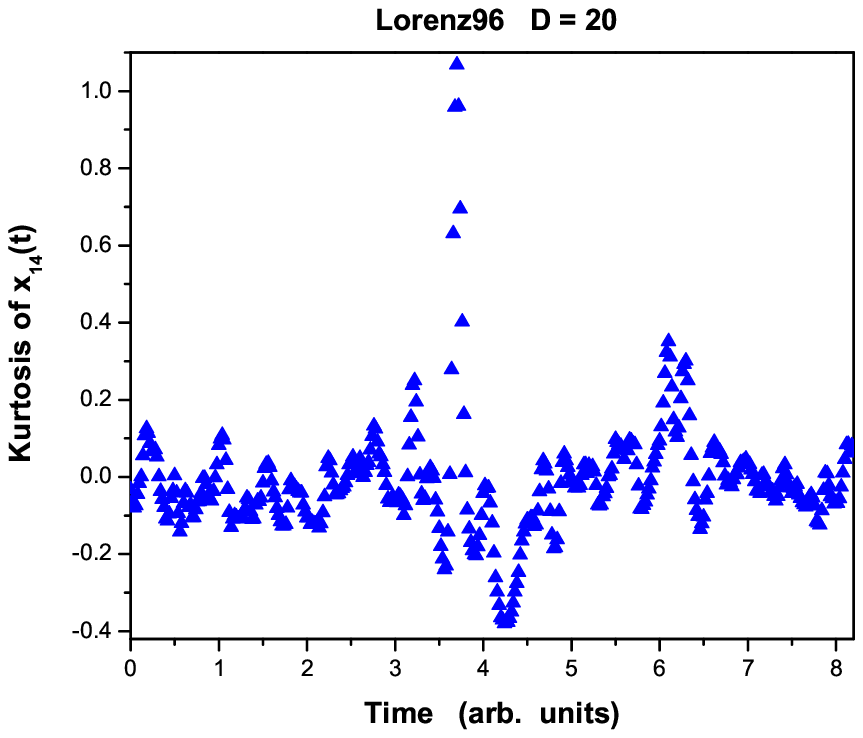}
\caption{{\bf Left} Skewness of the state variable $x_{14}(t)$, one of the unobserved quantities not presented to the model though the data assimilation path integral. {\bf Right} Kurtosis of the state variable $x_{14}(t)$, one of the unobserved quantities not presented to the model though the data assimilation path integral.}
\label{x14skewnessD20}
\end{figure}



\clearpage
\begin{thebibliography}{99}

\bibitem{abar2009} Abarbanel, H. D. I., ``Effective actions for statistical data assimilation,'' {\em Physics Letters A} {\bf 373},  4044–4048 (2009).

\bibitem{alexander05} Alexander, F. J., G. L. Eyink, and J. M. Restrepo,  ``Accelerated Monte Carlo
for optimal estimation of time series,'' {\em J. Stat. Phys.} {\bf 119},  1331-1345 (2005).

\bibitem{apte} Apte A, M. Hairer, A. Stuart, and J.  Voss, ``Sampling the posterior: An
approach to non-Gaussian data assimilation,'' {\em Physica D: Nonlinear
Phenomena,} {\bf 230}, 50–64 (2007). DOI: 10.1016/j.physd.2006.06.009.

\bibitem{restrepo} Restrepo, J. M., ``A path integral method for data assimilation,''
{\em Physica D: Nonlinear Phenomena,} {\bf 237}, 14–27. (2008). DOI:10.1016/j.physd.2007.07.020.




\bibitem{cox64} Cox, H., ``On the Estimation of State Variables and Parameters for Noisy Dynamic Systems,'' {\em IEEE Transactions on Automatic Control} {\bf AC-9}, 5-12 (1964).

\bibitem{fried66} Friedland, B. and I. Bernstein, ``Estimation of the State of a Nonlinear Process in the Presence of Nongaussian Noise and Disturbances,'' {\em J. Franklin. Inst.} {\bf 281}, 455-480 (1966).

\bibitem{jaz} Jazwinski, A. H., {\em Stochastic Processes and Filtering Theory}, Academic Press, 376 pp, (1970).

\bibitem{losch} Losch, M, R. Redler, and J. Schroter, ``Estimating a mean ocean state from hydrography and sea-surface height data with a nonlinear inverse section model: Twin experiments with a synthetic dataset,'' {\em J.. Oceano.} {\bf 32}, 2096-2112, (2002). 

\bibitem{lor96} Lorenz, E. N.,  ``Predictability – A problem partly solved,'' in {\em Proceedings
of the Seminar on Predictability, Volume 1}, ECMWF: Reading, UK, 1-18 (1996).

\bibitem{klwest} West, B. J. and Katja Lindenberg, {\em Nonequilibrium Statistical Mechanics of Open and Closed Systems}, 
VCH Publishers (1990), ISBN 047118683X and ISBN 9780471186830

\bibitem{mackay} MacKay, D. J.,  {\em Information theory, inference, and learning algorithms},
Cambridge University Press: Cambridge, UK (2003).

\bibitem{pap} Papoulis, A. {\em Probability, Random Variables, and Stochastic Processes, 2nd ed.} New York: McGraw-Hill, (1984). 

\bibitem{fano} Fano, R. M., {\em Transmission of Information: A Statistical Theory of Communications}, Wiley and MIT Press, (1961).

\bibitem{meister} DeWeese, M. R., and Markus Meister, ``How to measure the information gained from one symbol,'' {\em Network: Comput. Neural Syst.} {\bf 10}, 325–340 (1999). 



\bibitem{hamill} Hamill, T. M.,  {\em Predictability of weather and climate} in: T. Palmer, R. Hagedorn
(Eds.), Cambridge University Press, 124–156 (2006).

\bibitem{kost1} Abarbanel, H. D. I., M. Kostuk, and W. Whartenby, ``Data assimilation with regularized nonlinear instabilities,''
{\em Quarterly Journal of the Royal Meteorological Society} {\bf 136}, 769–783 (2010).

\bibitem{kost2} Kostuk, M., W. Whartenby, and H. D. I. Abarbanel, ``Determining the Number of Measurements Required for Nonlinear Data Assimilation,'' UCSD preprint, Winter, 2011.

\bibitem{quinn10} Quinn, J. C. and H. D. I. Abarbanel, ``State and parameter estimation using Monte Carlo evaluation
of path integrals,'' {\em Quarterly Journal of the Royal Meteorological Society} {\bf  136}, 1855-1867 DOI:10.1002/qj.690 (2010). 

\bibitem{neal} Neal, R. M., "Probabilistic Inference Using Markov Chain Monte Carlo Methods," Technical Report CRG-TR-93-1, Department of Computer Science, University of Toronto, 25 September 1993

\bibitem{hast} Hastings, W. K., "Monte Carlo Sampling Methods Using Markov Chains
and Their Applications," {\em Biometrika} {\bf 57}, 97-109 (1970). Stable URL: http://www.jstor.org/stable/2334940

\bibitem{feller} Feller, W., {\em An Introduction to Probability Theory and Its Applications} {\bf Volume 2}, John Wiley and Sons, New York, (1971).


\end{thebibliography}
\end{document}